\documentclass[fleqn,usenatbib]{mnras}

\usepackage[T1]{fontenc}
\usepackage{ae,aecompl}

\usepackage{graphicx}	
\usepackage{amsmath}	

\usepackage{amssymb}	
\usepackage{float}
\usepackage{caption}
\usepackage{subcaption}

\usepackage{ulem}
\usepackage{xcolor}

\usepackage{verbatim}

\title[VSFs of the hot ICM.]{Measuring the hot ICM velocity structure function using {\it XMM-Newton} observations} 

\author[Gatuzz et al.]{
Efrain Gatuzz$^{1}$\thanks{E-mail: egatuzz@mpe.mpg.de},
R. Mohapatra$^{2,3}$,
C. Federrath$^{3,4}$,
J. S. Sanders$^{1}$,
A. Liu$^{1}$,\newauthor
S. A. Walker$^{5}$,
C. Pinto$^{6}$
\\
$^{1}$ Max-Planck-Institut f\"ur extraterrestrische Physik, Gie{\ss}enbachstra{\ss}e 1, 85748 Garching, Germany\\
$^{2}$ Department of Astrophysical Sciences, Princeton University, NJ 08544, USA \\
$^{3}$ Research School of Astronomy and Astrophysics, Australian National University, Canberra, ACT 2611, Australia\\
$^{4}$ Australian Research Council Centre of Excellence in All Sky Astrophysics (ASRO3D), Canberra, ACT 2611, Australia\\ 
$^{5}$ Department of Physics and Astronomy, University of Alabama in Huntsville, Huntsville, AL 35899, USA\\
$^{6}$ INAF - IASF Palermo, Via U. La Malfa 153, I-90146 Palermo, Italy 
}

\date{Accepted XXX. Received YYY; in original form ZZZ} 
\pubyear{2023} 
\begin{document}
 \label{firstpage}
\pagerange{\pageref{firstpage}--\pageref{lastpage}}
\maketitle 

\begin{abstract}
It has been shown that the gas velocities within the intracluster medium (ICM) can be measured by applying novel {\it XMM-Newton} EPIC-pn energy scale calibration, which uses instrumental Cu K$\alpha$ as reference for the line emission. Using this technique, we have measured the velocity distribution of the ICM for clusters involving AGN feedback and sloshing of the plasma within the gravitational well (Virgo and Centaurus) and a relaxed one (Ophiuchus). We present a detailed study of the kinematics of the hot ICM for these systems. First, we compute the velocity probability distribution functions (PDFs) from the velocity maps. We find that for all sources the PDF follows a normal distribution, with a hint for a multimodal distribution in the case of Ophiuchus. Then, we compute the velocity structure function (VSF) for all sources in order to study the variation with scale as well as the nature of turbulence in the ICM. We measure a turbulence driving scale of  $\sim 10-20$~kpc for the Virgo 
cluster,  while the Ophiuchus cluster VSF reflects the absence of strong interaction between the ICM and a powerful Active Galactic Nucleus (AGN) at such spatial scales. For the former, we compute a dissipation time larger than the jet activity cycle, thus indicating that a more efficient heating process than turbulence is required to reach equilibrium. This is the first time that the VSF of the hot ICM has been computed using direct velocity measurements from X-ray astronomical observations. 
\end{abstract}



\section{Introduction}\label{sec_in}   
Measuring the velocity structure of the ICM is important in order to constrain the different heating mechanisms that have been proposed to transfer energy from active galactic nuclei (AGN) back into the ICM \citep[see][for a review]{fab12b}. In addition to energetics, turbulent motions also contribute to non-thermal pressure support, particularly at large radii and affect cluster mass estimates when assuming hydrostatic equilibrium \citep[e.g. ][]{lau09,eck19}. They play a role in the transport of metals within the ICM, due to uplift and sloshing of metals by AGN outflows \citep[e.g.][]{sim08,wer10}. The turbulent velocity structure is also an excellent probe of the microphysics of the ICM, such as viscosity and conductivity \citep{gas14,Zhu19}. In addition, measuring velocities should directly measure the sloshing of gas in cold fronts, which can remain for several Gyr \citep{roe12,roed13,wal18}.

Simulations indicate that the ICM should contain turbulent and bulk flow motions, due to the merger of other subcomponents and clusters \citep{lau09,vaz11,sch17,haj18,lim18,vaz21}. Inflation of bubbles and the action of the relativistic jets by the central AGN also likely generate motions of a few hundred km/s  \citep{bru05,hei10,ran15,yan16,bam19}. Furthermore, merging substructures can generate relative bulk motions of several hundred km/s due to perturbations in the ICM \citep{asc06,ich19,vaz18,zuh18}. Overall, there is a close connection between the ICM physical state and the velocity power spectra \citep{gas14}. 

Despite its importance, the velocity structure of the ICM remains poorly constrained observationally. Direct measurements of random and bulk motions in the ICM using Fe-K emission lines were obtained by the {\it Hitomi} observatory, revealing low levels of turbulence near the Perseus cluster core despite the obvious impact of the AGN and its jets on the surrounding ICM \citep{hit16}. \citet{ota16} examined several clusters with {\it Suzaku}, although systematic errors from the {\it Suzaku} calibration were likely around $300$ km/s and its PSF was large. Low turbulence motion is also measured from line broadening and resonant scattering, with velocities between $100-300$ km/s and limited to the cluster core  \citep{san10,san13,pin15,ogo17,liu19}. Indirect estimates of the level of turbulent velocities have been obtained from X-ray brightness fluctuations \citep{zhu14,zhu18} and thermal Sunyaev-Zeldovich fluctuations \citep{zel69,sun70,kha16}. However, these methods are not based on direct velocity measurements and are highly model-dependent.

\citet{san20} present a novel technique that consists of using instrumental X-ray lines seen in the spectra of the {\it XMM-Newton} EPIC-pn detector to calibrate the absolute energy scale of the detector to better than $100$ km/s at Fe-K. Using this technique, direct measurements of the bulk ICM velocity distribution have been done in multiple systems, including the Perseus and Coma clusters \citep{san20}, the Virgo cluster \citep{gat22a,gat22d}, the Centaurus cluster \citep{gat22b} and the Ophiuchus cluster \citep{gat22c}.

Velocity structure functions ($\mathrm{VSF}$s) and spatial power spectra constitute useful diagnostic tools to study the turbulence motions in a medium (e.g., ISM or ICM), since they represent the variation of velocity with scale \citep{fed10,fed21,set23}. Recent observational studies have used such diagnostic to study the interstellar medium \citep[e.g. ][]{xu20a,ha21,mar21,che23} and intergalactic medium \citep[e.g. ][]{xu20b} velocity structure. \citet{li20a} studied turbulent velocities of ICM using optical data of atomic filaments in several nearby clusters. They analyzed the first-order structure functions of line-of-sights (LOS) velocity ($\mathrm{VSF}_1^{\mathrm{LOS,obs}}$) and found them to be steeper than expected from Kolmogorov turbulence theory \citep{kol41}. They also found that the driving scale of turbulence in their sample of clusters is proportional to the size of AGN-driven bubbles. Such measurements have led to numerical studies of $\mathrm{VSF}$s and velocity power spectra in similar multiphase ICM environments \citep[e.g.][]{zuh16b,moh19,hil20,nel20,wan21}. More recently, \citet{moh22} carried out a thorough study of the VSFs for both the hot and cold ICM phases, including the effect of projection using different weightings along the LOS.

In this work, we study the nature of the ICM within the Virgo, Centaurus and Ophiuchus clusters by measuring their VSF using direct velocity measurements obtained with the {\it XMM-Newton} EPIC-pn detector. The outline of this paper is as follows. In Section~\ref{sec_dat} we describe the data reduction and fitting process.  In Section \ref{sec_velfun} we analyze the velocity probability distribution functions. The analysis of the VSFs is shown in Section~\ref{first_vsf}. A detailed discussion of the results is shown in Section~\ref{sec_dis}, while the conclusions and summary are included in Section~\ref{sec_con}. Throughout this paper we assumed a $\Lambda$CDM cosmology with $\Omega_m = 0.3$, $\Omega_\Lambda = 0.7$, and $H_{0} = 70 \textrm{ km s}^{-1}\ \textrm{Mpc}^{-1} $.

\section{XMM-Newton data reduction}\label{sec_dat}
The {\it XMM-Newton} European Photon Imaging Camera \citep[EPIC,][]{str01} observations are the same as we used in \citet{gat22a,gat22b,gat22c} and we followed the same data reduction process. Spectra were reduced with the Science Analysis System (SAS\footnote{\url{https://www.cosmos.esa.int/web/xmm-newton/sas}}, version 19.1.0). First, we processed each observation with the {\tt epchain} SAS tool. We used only single-pixel events (PATTERN==0) while bad time intervals were filtered from flares applying a 1.0 cts/s rate threshold. In order to avoid bad pixels or regions close to CCD edges we filtered the data using FLAG==0. 

Following the work done in \citet{san20,gat22a, gat22b, gat22c}, we used updated calibration files, which allows to obtain velocity measurements down to 100~km/s at Fe-K by using the background X-ray lines identified in the spectra of the detector as references for the absolute energy scale. Identification of point sources was performed using the SAS task {\tt edetect\_chain}, with a likelihood parameter {\tt det\_ml} $> 10$. The point sources were excluded from the subsequent analysis, including the AGN in the Virgo cluster core (i.e., a central circular region with a diameter $D=58$~\arcsec).

We made spectral maps of the clusters using the contour binning algorithm \citep{san06}. We created regions applying a geometrical constraint factor of 1.7, to prevent bins becoming too elongated. We masked out the point sources during binning. We performed the analysis using the maps binned with a signal-to-noise ratio of 75. While these maps have the disadvantage of producing a non-smoothly varying map, compared to those analyzed in \citet{gat22a,gat22b,gat22c}, the advantage is that the regions are statistically independent. We analyze the spectra with the {\it xspec} spectral fitting package (version 12.11.1\footnote{\url{https://heasarc.gsfc.nasa.gov/xanadu/xspec/}}) using {\tt cash} statistics \citep{cas79}.

For each source, we followed the spectral fitting described in \citet{gat22a,gat22b,gat22c} which we will describe briefly. We model the cluster gas emission with an {\tt apec} model. In the case of Centaurus, we model the spectra with a log-distribution of temperatures ({\tt lognorm} model) in order to account for the ICM multi-temperature component within the system \citep{gat22b}. In order to account for the Galactic absorption we included a {\tt tbabs} component \citep{wil00}. The free parameters in the model are the redshift, metallicity, temperature, log$\sigma$ (i.e., for the {\tt lognorm} model) and normalization. Finally, we included Cu-$K\alpha$, Cu-$K\beta$, Ni-$K\alpha$ and Zn-$K\alpha$ emission lines to model the instrumental background.

\begin{figure}    
\centering
\includegraphics[width=0.47\textwidth]{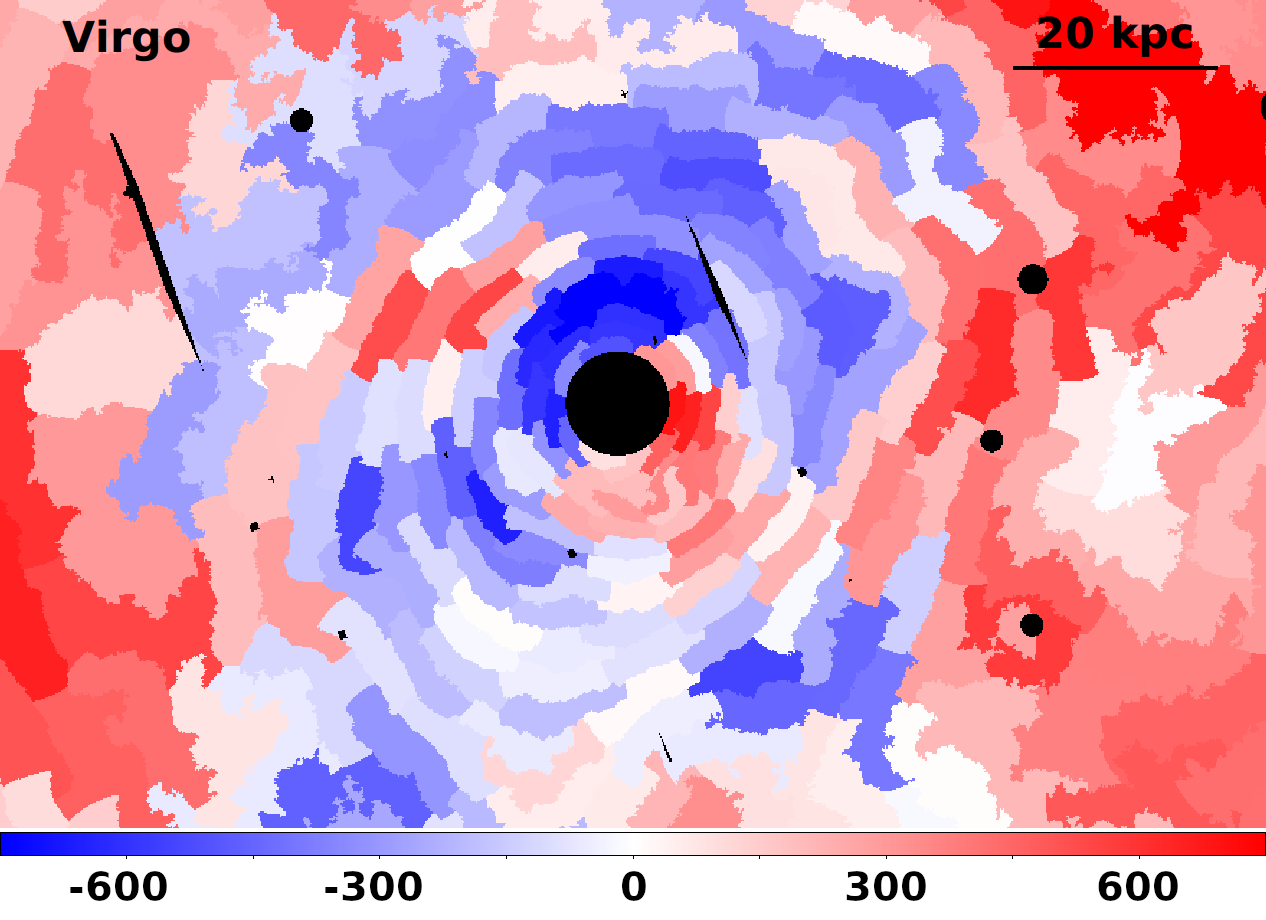}
\includegraphics[width=0.47\textwidth]{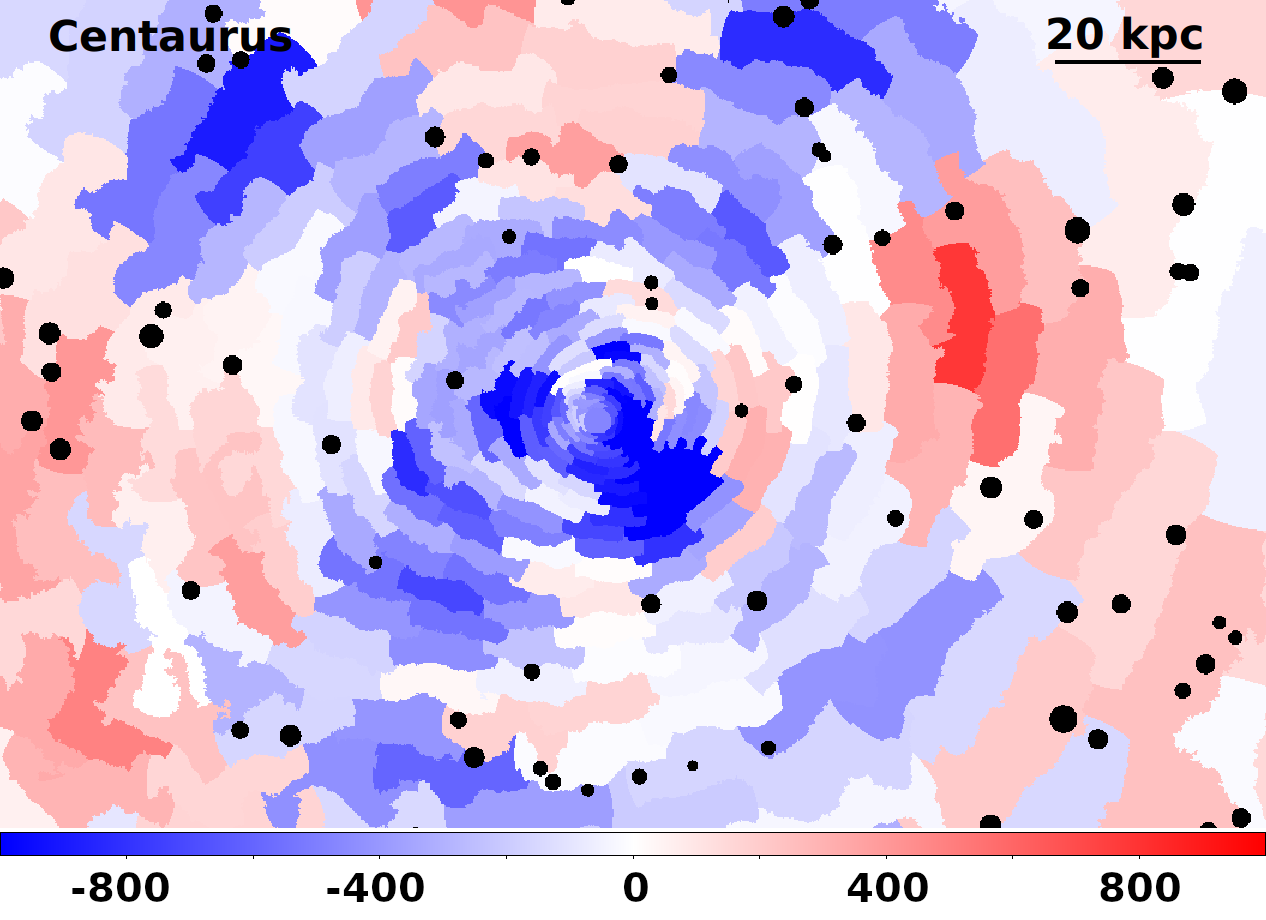}
\includegraphics[width=0.47\textwidth]{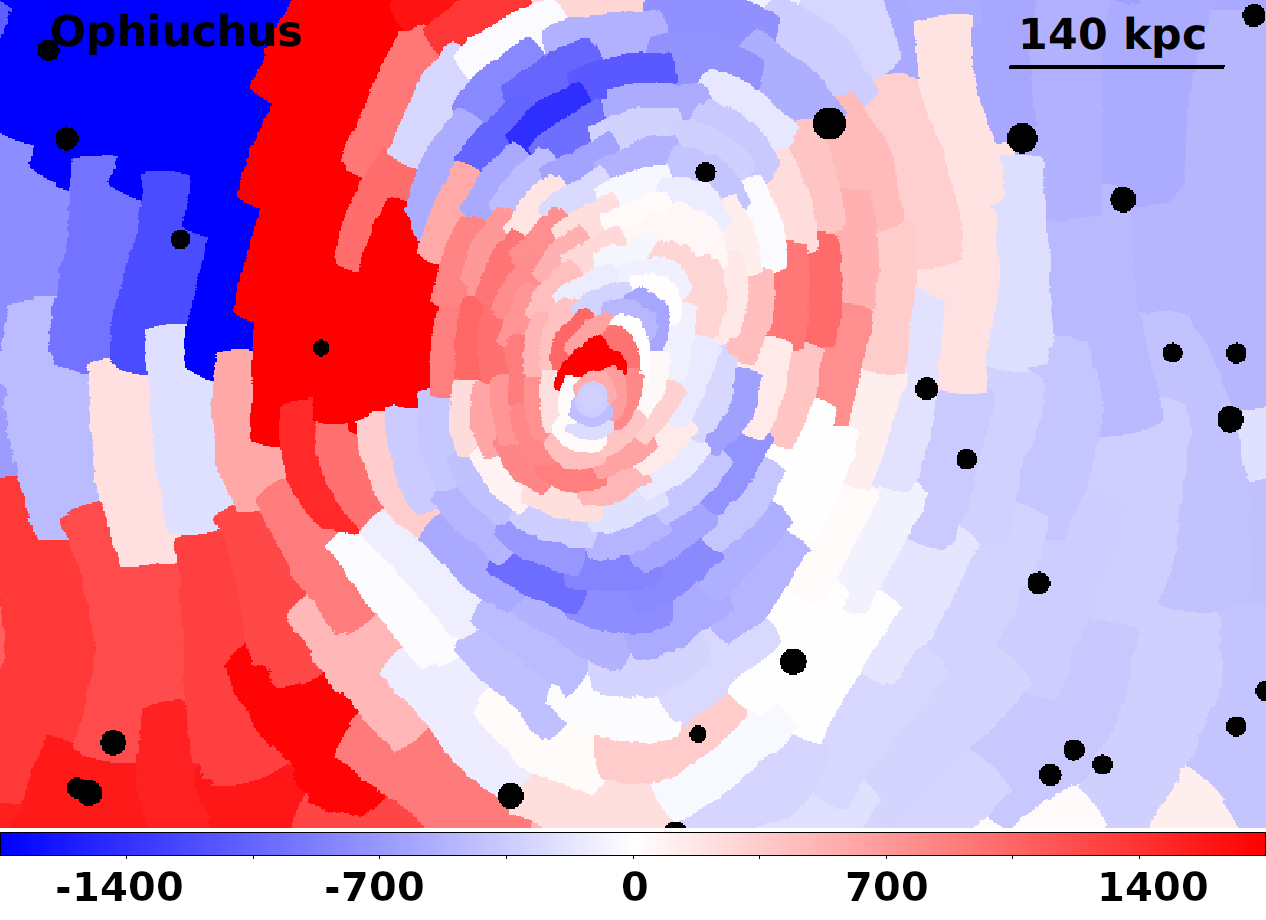} 
\caption{Spatially resolved velocity map (km/s) of the hot ICM created using the contour binning method for the Virgo (top panel), Centaurus (middle panel) and Ophiuchus cluster (bottom panel). The maps were created by fitting the spectra extracted from regions with a signal to noise ratio of 75.} \label{fig_contbin_vel} 
\end{figure}

Figure~\ref{fig_contbin_vel} shows the resulting velocity map from the contour binning process. The line shifts are with respect to each system analyzed (i.e., not with respect to us). Overall, these maps are similar to the codependent velocity maps described in our previous reports. For example, the Virgo cluster displays a redshifted gas along the west direction near the cluster center, while a blueshifted gas along the east direction is seen, a distribution shown in \citet[][in Figure~11]{gat22a}. The Centaurus cluster, on the other hand, shows mainly a blueshifted gas, with larger velocities around the south-west direction, similar to the structure found in \citet[][in Figure~5]{gat22b}. Finally, a redshifted-to-blueshifted interface with very large velocities can be identified in the Ophiuchus cluster velocity map in the east direction from the central core, a feature that was also identified in \citet[][in Figure~9]{gat22c}.

\section{Velocity probability distribution functions}\label{sec_velfun} 
\begin{figure}    
\centering
\includegraphics[width=0.47\textwidth]{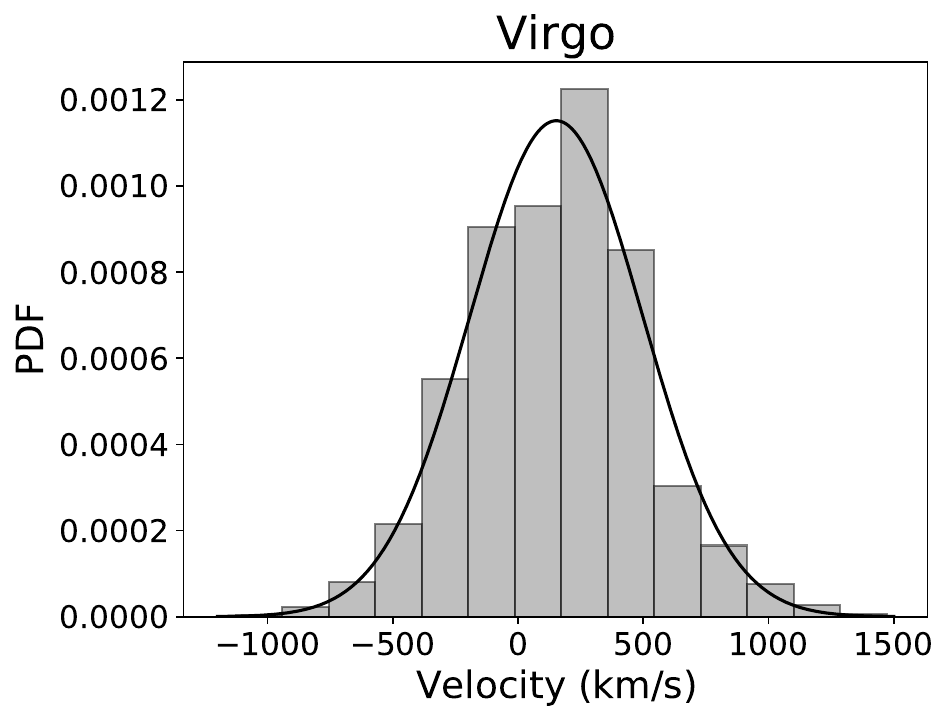}\\
\includegraphics[width=0.47\textwidth]{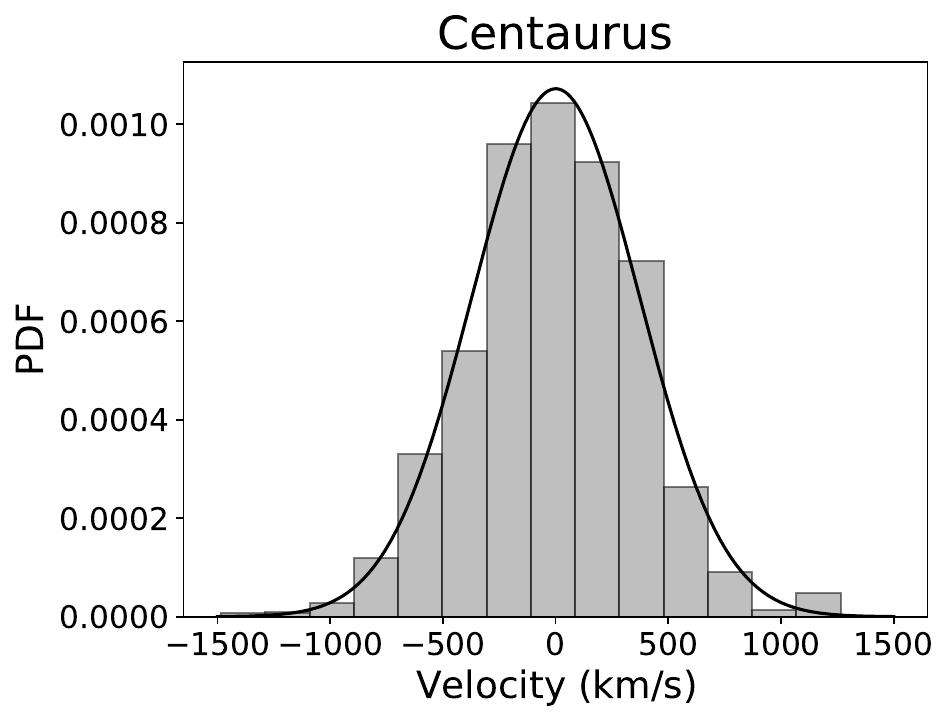}\\
\includegraphics[width=0.47\textwidth]{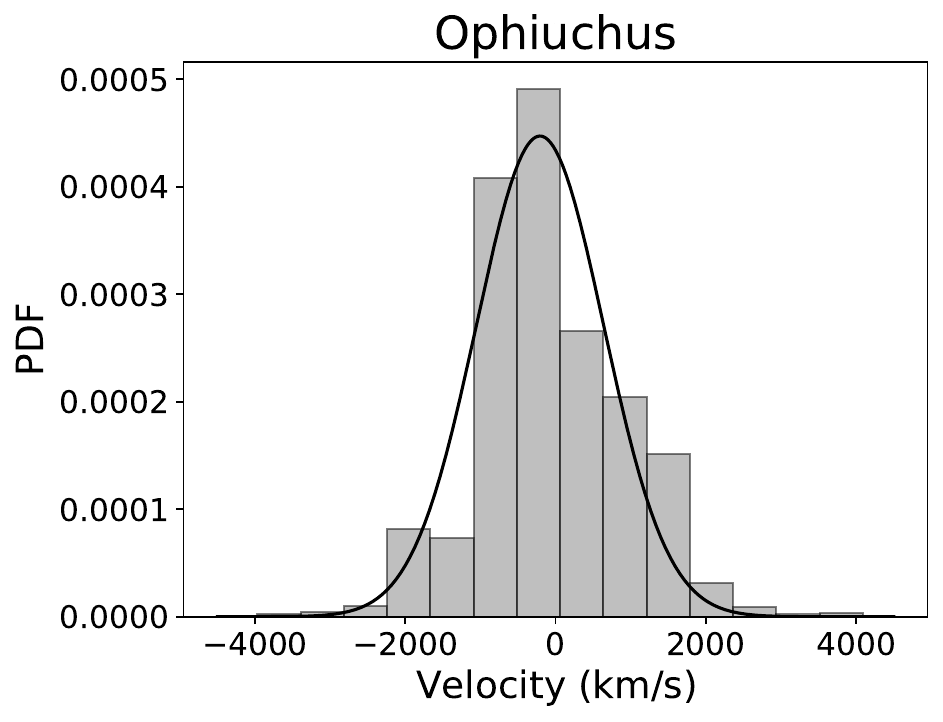} 
\caption{Velocity PDFs of the hot ICM obtained for the Virgo (top panel), Centaurus (middle panel) and Ophiuchus (bottom panel) velocity maps. The black line corresponds to the best-fit using a Gaussian model.}\label{fig_PDF_vel} 
\end{figure}

Figure \ref{fig_PDF_vel} shows the velocity probability distribution functions (PDFs) computed from the velocity maps, weighted by area. For each PDF, we compute the Shapiro-Wilk \citep{sha65} and D'Agostino and Pearson's \citep{dag73} normality tests to determine if the data set is well modeled by a Gaussian\footnote{Both tests are included in the {\tt scipy.stats} package.}. We found that for all sources the distribution follows a normal distribution (i.e., the $p$-value is larger than $\alpha=0.05$ level). Table~\ref{tab_pdf_gauss} shows the best-fit parameters obtained for a Gaussian model fitted to each PDF. 

In the cases of Ophiuchus, there are hints for a multimodal probability distribution function, however the sample of points for the modes is not large enough to perform a normality test. Simulations predict a Gaussian velocity probability distribution function for the ICM as the one observed in Figure~\ref{fig_PDF_vel} \citep{zuh16,ehl21,wan21}.  \citet{moh19,moh22} predicted a $\sigma$ value in the $\sim 200-400$~km/s range for the hot ICM phase, with indications of isotropy, while {\it Hitomi} measured up to $\sigma\sim220$~km/s. However, these measurements were obtained only for the Perseus cluster. 

Large velocities can be observed due to the presence of substructures, subgroups and/or strong merger shocks \citep{ang15,ang16,ota16}. \citet{san20} found large velocities ($>1000$ km/s) for some regions within the Perseus cluster and the Coma cluster. In the latter case, they are associated with subgroups in the system. For Ophiuchus cluster, the large velocities found along the east direction from the cluster center (see Figure~\ref{fig_contbin_vel}), coincide with a sharp surface brightness, which can be an indication of merger activity \citep{gat22c}.  A detailed analysis of possible systematics carried out by \citet[][see Section 3.2]{gat22c} lead out to the conclusion that these velocity patterns are significant and reliable. It could be that the merger direction within this source has a large line-of-sight component, therefore two separated clumps of gas are not observed in the image. Previous results obtained from X-ray, radio and optical observations are consistent with Ophiuchus being a merger \citep{wat01,mur09,mil10,ham12}. Also, large velocities have been found using optical observations for cluster members even for distances $<150$~kpc \citep{dur15}. Future missions such as the X-ray Imaging and Spectroscopy Mission \citep[XRISM,][]{xri20}, the Line Emission Mapper \citep[LEM,][]{lem22} or Athena \citep{nan13} will provide more direct evidence to test such interpretation.

\begin{table}
\caption{\label{tab_pdf_gauss}Velocity PDFs best-fit parameters.}
\centering
\begin{tabular}{ccccccc}
\\
Source &\multicolumn{3}{c}{{\tt gaussian}} & $\chi^{2}$/dof \\
\hline
 &$\sigma$ & $\mu$&  $norm$  & \\  
Virgo&$344\pm 18$& $30\pm 19 $& $0.00115\pm 0.00005 $&$1.06$\\
Centaurus&$371\pm 12 $&$-15\pm 12 $& $0.00107\pm 0.00003$& $1.08$\\
Ophiuchus &$847\pm 92 $& $-207\pm 93$& $0.00044\pm 0.00004 $&$1.09$\\ 
\\ 
 \hline
\multicolumn{5}{l}{$\sigma$ and $\mu$ in units of km/s.}\\
\end{tabular}
\end{table}

\section{Velocity structure functions}\label{first_vsf}
\begin{figure}    
\centering
\includegraphics[width=0.48\textwidth]{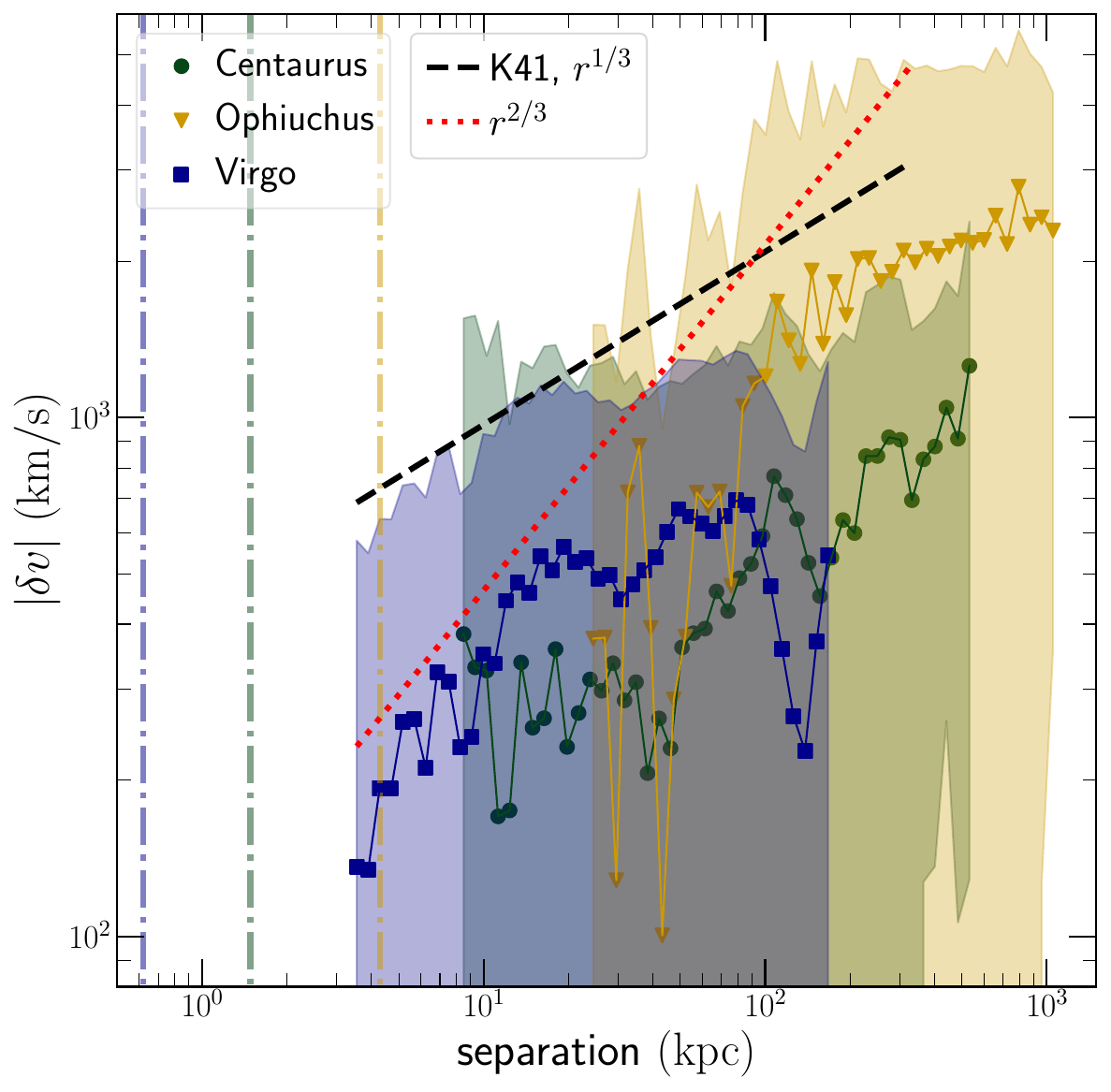} 
\caption{VSFs of the hot ICM obtained for Virgo (blue points), Centaurus (green points) and Ophiuchus (yellow points) galaxy clusters. Power laws with slope $\sim$1/3 (i.e., Kolmogorov) and $\sim$2/3 are also included. The shaded regions show the $1-\sigma$ interval around the mean VSF. The vertical `dashdot' lines indicate the PSF scale.} \label{fig_vsfs} 
\end{figure}

We compute the first-order structure function by taking the weighted average of the difference between the line-of-sight velocities ($v$) of two points separated by $r$. 
Mathematically, we define it as:
\begin{equation}
    \delta v (r) = \frac{\sum_{\mathbf{x}}{w(\mathbf{x}+\mathbf{e_1}r, \mathbf{x}) |v(\mathbf{x}+\mathbf{e_1}r)-v(\mathbf{x})|}}{\sum_\mathbf{x}{w(\mathbf{x}+\mathbf{e_1}r, \mathbf{x})}}
\end{equation}
where $x$ denotes the position of any point in the dataset and $\mathbf{e_1}$ denotes a unit-vector in any direction. We bin $\delta v$ into logarithmically-spaced bins of separation $r$. We have used three different weighting functions for our analysis:
\begin{subequations}
\begin{align}
    &w_{\mathrm{const}}=\mathrm{const},\\
    &w_{\mathrm{area}}=\mathrm{area}_{\mathbf{x}+\mathbf{e_1}r}+\mathrm{area}_{\mathbf{x}},\\
    &w_{\mathrm{err}}=\left[\left(v_{\mathrm{err}}\right)^2_{\mathbf{x}+\mathbf{e_1}r}+\left(v_{\mathrm{err}}\right)^2_{\mathbf{x}}\right]^{-1/2}.
\end{align}
\end{subequations}
Here area denotes the total number of pixels in a binned region (see Figure~\ref{fig_contbin_vel}). We use $w_{\mathrm{const}}$ for the data presented in Figures~\ref{fig_vsfs}, \ref{fig_vsf_virgo_hc} and \ref{fig_vsf_centaurus_hc}. We show the effects of choosing different weighting functions in Figure~\ref{fig_vsfs_systematic}.

Figure~\ref{fig_vsfs} shows the first-order VSFs computed for Virgo (blue points), Centaurus (green points) and Ophiuchus (yellow points). Because of the large uncertainties, we limit the analysis to the first-order VSF. A broad power-law slope can be identified for all sources (red dotted line), confirming that the gas motion is turbulent. While a slope of the VSF $\sim$1/3 is consistent with the expectation of classical Kolmogorov turbulence for an incompressible fluid, it has been shown that the steepening of the VSF may be due to projection effects \citep{moh22}\footnote{Using idealised turbulence simulations, \cite{moh22} show that projection along the line-of-sight generally leads to steepening of the VSF. However, the degree of steepening decreases with increasing clumpiness of the emitting medium, see their Fig.~7. In the absence of a systematic study on the steepening of the VSF with clumpiness, we refrain from guessing the true 3D slope from the values obtained for the projected slope.}. For Virgo cluster a flattening is observed, thus indicating a driving scale of $\sim$15~kpc. Such flattening has been shown in magnetohydrodynamical simulations of AGN jet feedback \citep{wan21,moh22}.   
In the case of Ophiuchus, such flattening is not observed. This is expected given that the influence of AGN feedback is minimal for this system. There are hints of flattening on very large scales ($\sim 250)$~kpc. \citep{gia20} reported the discovery of a very large bubble of radius $\sim 230$~kpc. However, the analysis of systematic effects shows that such flattening may be artificial (see Section~\ref{sys_effec}). Finally, the smallest scale accessible in our analysis is limited by the Full Width at Half Maximum (FWHM) of the effective point-spread-function (PSF). Recent work by \citet{che23} suggests that the steepening of the VSF could also partially be due to the total PSF, which tends to smooth out velocity differences on turbulence-driving scales much smaller than the scales we have derived in our study. Additionally, we require a minimum of 1000 counts in the 4-9 keV energy range to accurately determine the redshift of the Fe K-complex. Consequently, we are extracting spectra from regions considerably larger than the PSF.

\section{Discussion}\label{sec_dis}

The VSFs indicate a driving scale of turbulence of $\sim 10-20$~kpc for the Virgo cluster. 
Moreover, such a driving scale is expected given that bubbles have typical sizes of $\sim 5-20$~kpc \citep{zhu14}. We can estimate the dissipation time, which is a few times the eddy turnover time $t_{\ell}\simeq l/v_{\ell}$, where $\ell$ is the scale and $v_{\ell}$ is the velocity at that scale. For the Virgo 
we take $v_{\ell}\simeq 350$~km/s, which gives $t_{\ell}(10$~kpc$)\approx 28$~Myr and a dissipation time of $t_{\mathrm{diss}}(10$~kpc$)>40$~Myr. The period of AGN outburst is $t_{\mathrm{AGN}}\approx 12$~Myr for the Virgo cluster \citep{for17} .
Thus, the dissipation time is longer than the jet activity cycle, therefore the turbulence can transfer only a small fraction of the AGN power to heat the ICM. This implies that more efficient heating processes in addition to turbulence are required to reach equilibrium \citep[e.g., ICM mixing with hot bubbles][]{hil20}. However, we note that in both cases $t_{\ell}$ and $t_{\mathrm{AGN}}$ are highly uncertain.

The driving scale obtained for Virgo is larger than that obtained for the cold gas by \citet[][$\sim1-2$~kpc]{li20a} near the cluster core. In that sense, \citet{moh22} have shown that the cold- and hot-phase velocities are uncorrelated at scales close to the driving scale. In the case of the Ophiuchus cluster the AGN itself only displays weak, point-like radio emission. While \citet{gia20} report to have discovered a large cavity to the southeast of the cluster, \citet{gat22c} did not find changes in the metallicities or temperatures for regions inside and outside that region.
 
\begin{figure}    
\centering
\includegraphics[width=0.48\textwidth]{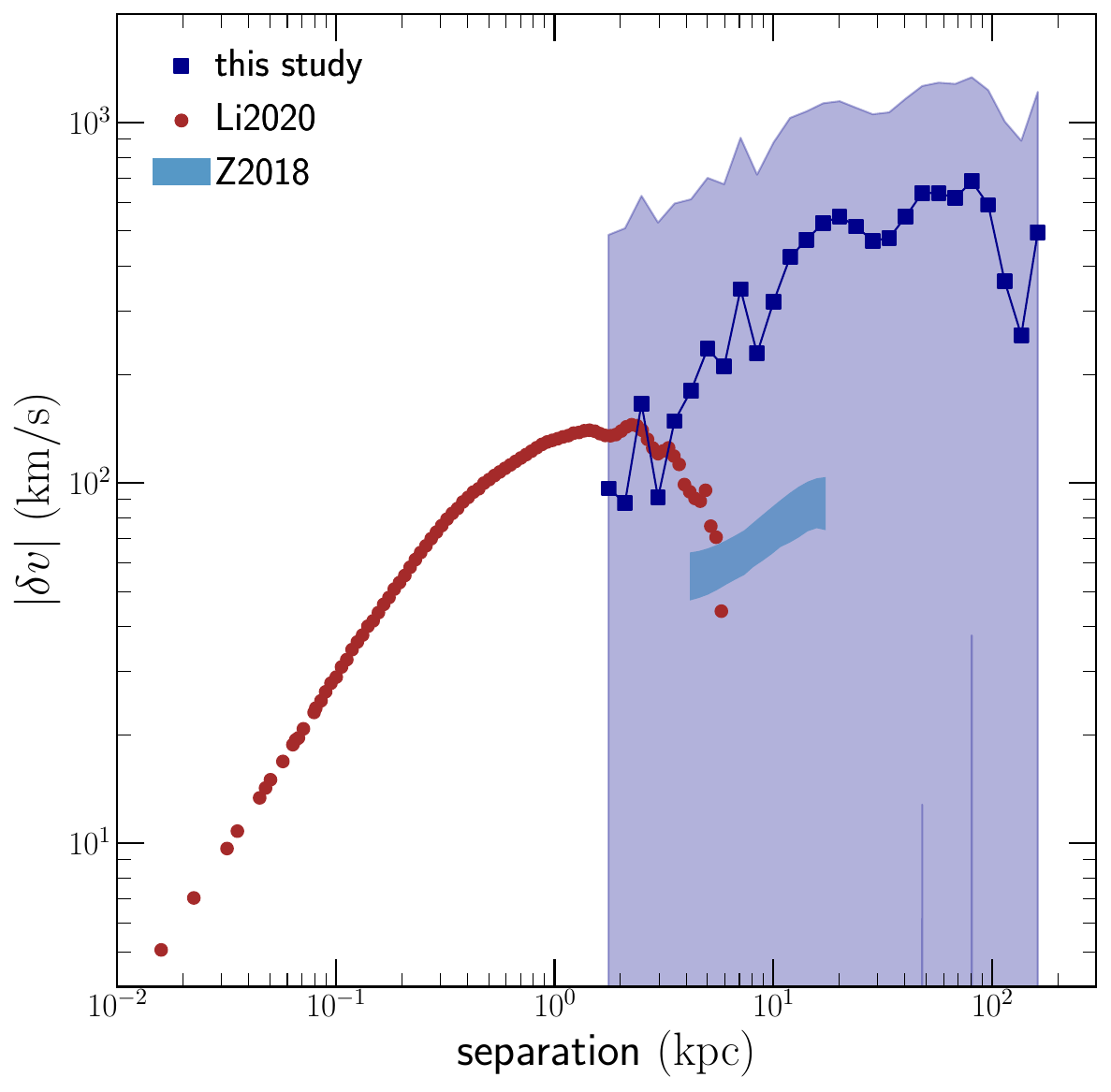} 
\caption{VSFs of the hot ICM obtained for Virgo (blue squares) in this study, compared with the cold ICM measured by \citet{li20a} (brown circles) using MUSE data of H$\alpha$ filaments and \citet{zhu18} (blue shaded region) using velocities inferred from X-ray brightness fluctuations in Chandra observations.} \label{fig_vsf_virgo_hc} 
\end{figure}

\begin{figure}    
\centering
\includegraphics[width=0.48\textwidth]{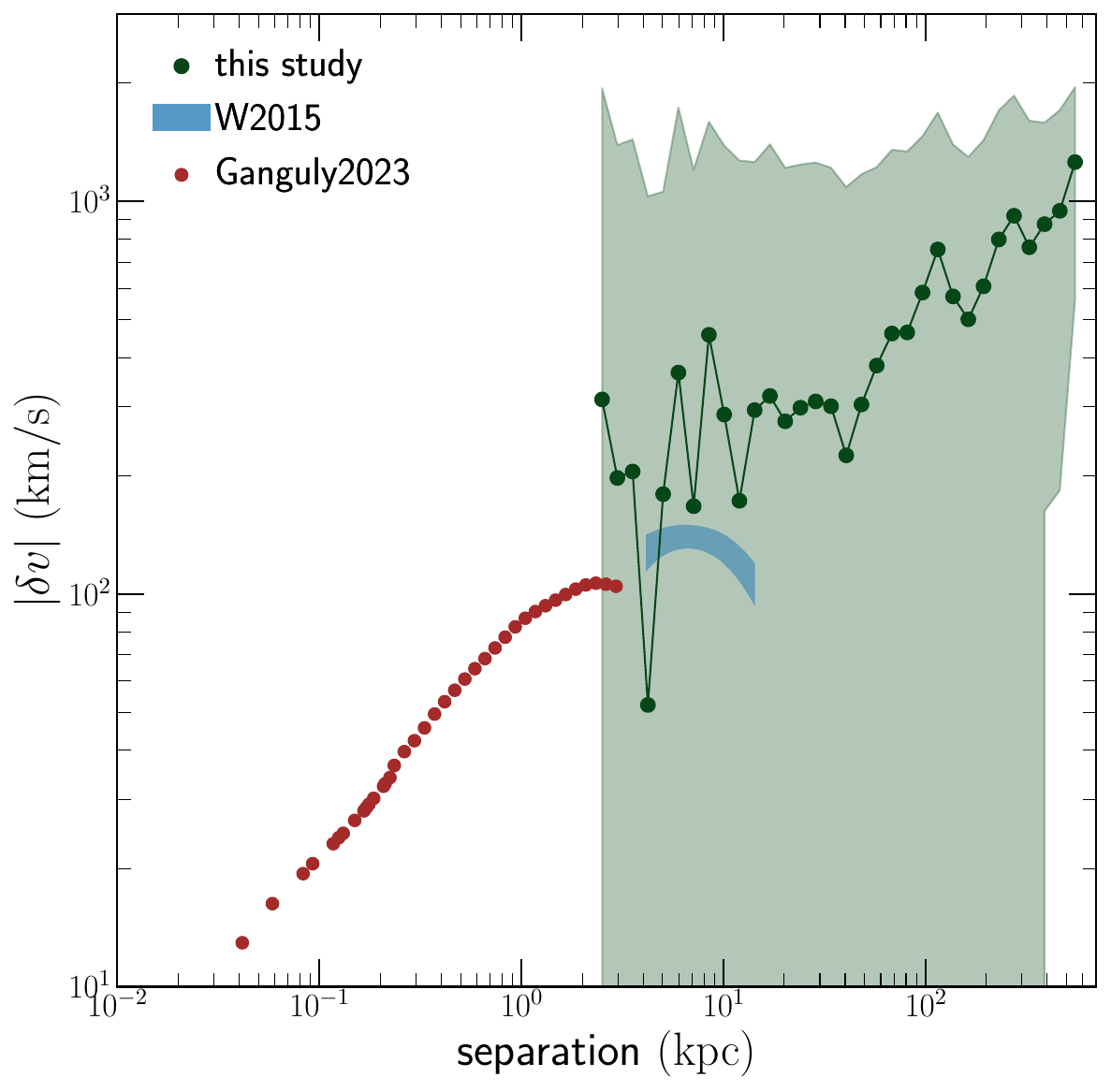} 
\caption{VSFs of the hot ICM obtained for Centaurus (green circles) in this study, compared with \citet{wal15} (blue shaded region) using velocities inferred from X-ray brightness fluctuations in Chandra observations.} \label{fig_vsf_centaurus_hc} 
\end{figure}

Figures~\ref{fig_vsf_virgo_hc} and \ref{fig_vsf_centaurus_hc} show comparisons between the VSF of the hot ICM obtained for Virgo and Centaurus (red circles) from \citet{li20a} and \citet{gang23}, respectively, using MUSE data of H$\alpha$ filaments (i.e., the cold ICM). The H$\alpha$ velocities inferred on the largest scale seem to match the velocities inferred from the X-ray observations on small scales. This may imply multiple or different driving scales for the hot and cold gas, since the flattening occurs at different separations ($\sim$2~kpc for the cold gas and $\sim$10~kpc for the hot gas). However, the smaller field of view of the observations analyzed by \citet{li20a} could also affect the overall shape of the curve on large scales (a few kpc). In that sense, future observations are crucial to better understand the link between both environments. For example, H$\alpha$ measurements on larger scales may show additional energy injection scales while better high-resolution X-ray VSFs will provide insights about additional breaks on smaller scales. 
 
 We also compare our measurements with the velocities inferred from X-ray brightness fluctuations in \citet{zhu18} for Virgo and \cite{wal15} for Centaurus in Figures~\ref{fig_vsf_virgo_hc} and \ref{fig_vsf_centaurus_hc}, resepectively. We find that the two measurements differ by roughly a factor of $2$--$4$ (note the large errors in our data on scales $\lesssim10~\mathrm{kpc}$, as well as our measurement uncertainty of $\sim100~\mathrm{km/s}$). In addition to the above, further differences could be due to the following reasons: (1) The region analyzed by \citet{zhu18} is very small in comparison with our analysis; (2) In the case of Virgo, unlike us, \cite{zhu18} exclude the jet-arm regions from their calculations \citep[see extended data fig.~2 in][]{zhu14}, which are associated with larger brightness (and possibly velocity) fluctuations; (3) stratified turbulence simulations have shown that the ratio between density and velocity fluctuations that they use depends on the strength of stratification of the ICM \citep{Moh21}. It increases with increasing stratification and saturates for strongly stratified turbulence. Since both \cite{wal15} and \cite{zhu18} use the value of this ratio in the limit of strong stratification \citep[see Fig.~10 in][]{Moh21}, they may under-estimate the amplitude of turbulent velocities when the stratification is weaker.  In that sense, recent works have estimated the expected scatter for the proportionality coefficient  \citep[see for example Figure 6 in ][]{zhu23}.

\subsection{Systematic effects}\label{sys_effec}
\subsubsection{Effect of weighting function}\label{weight_function}
\begin{figure}    
\centering
\includegraphics[width=0.47\textwidth]{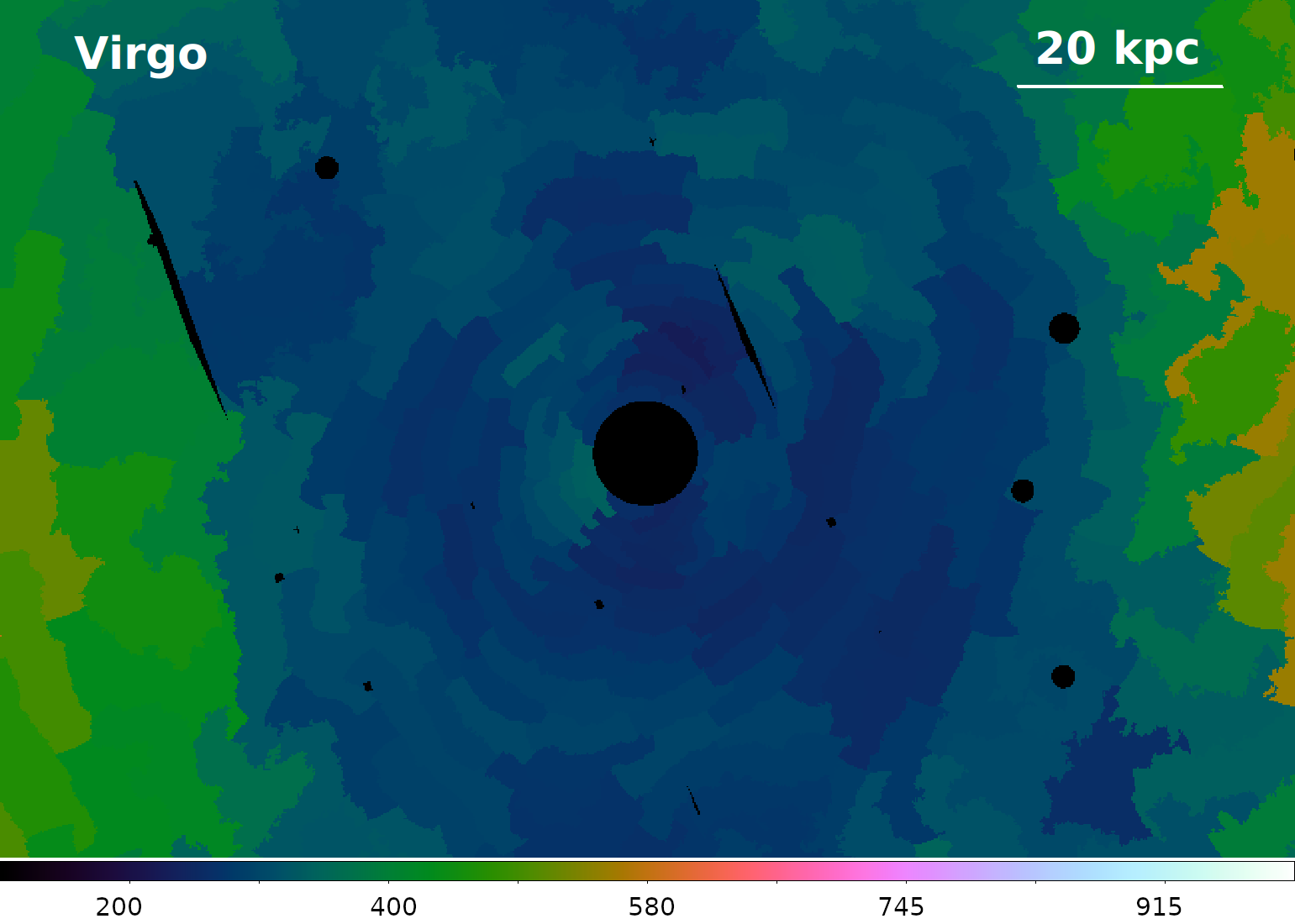}
\includegraphics[width=0.47\textwidth]{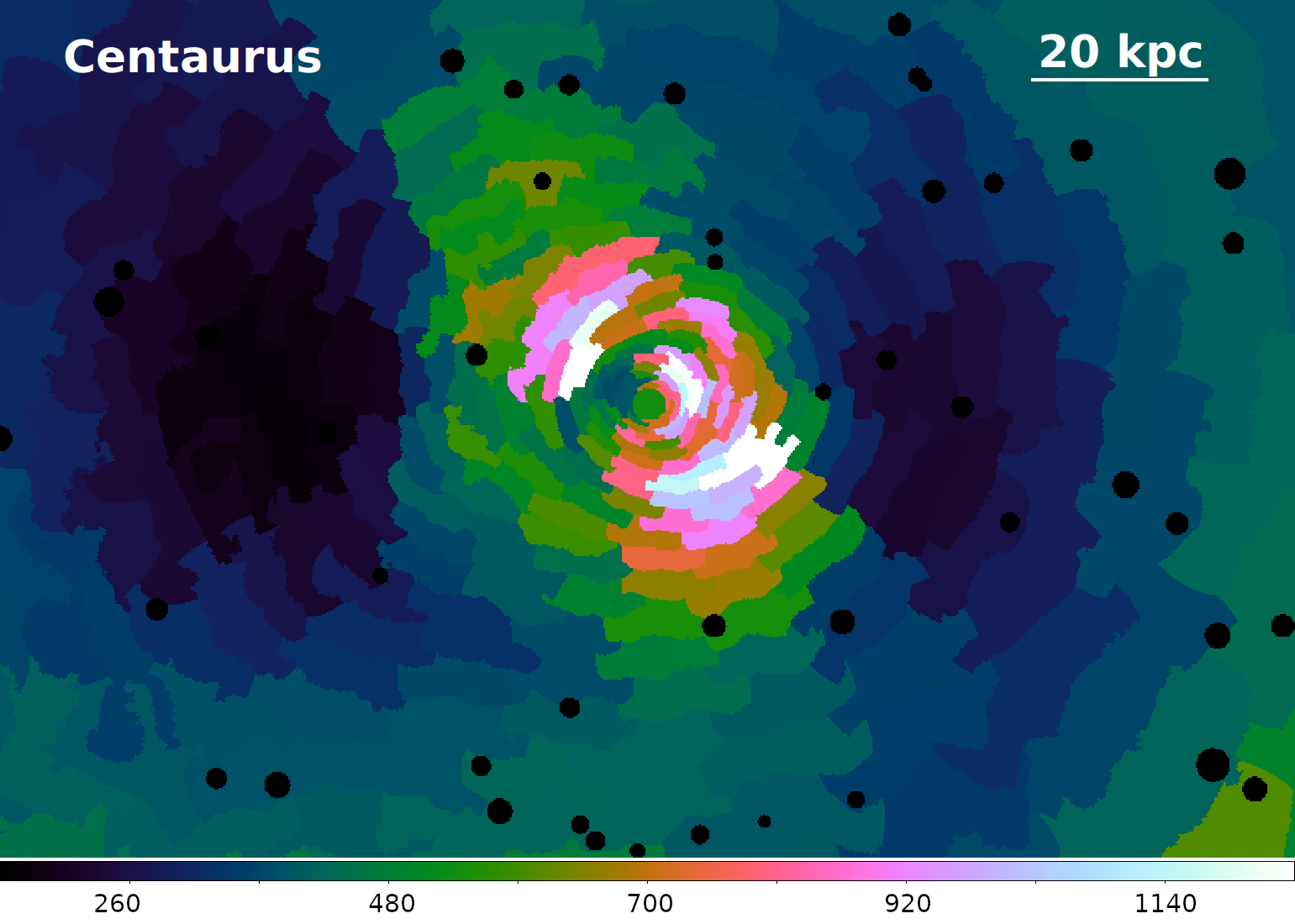}
\includegraphics[width=0.47\textwidth]{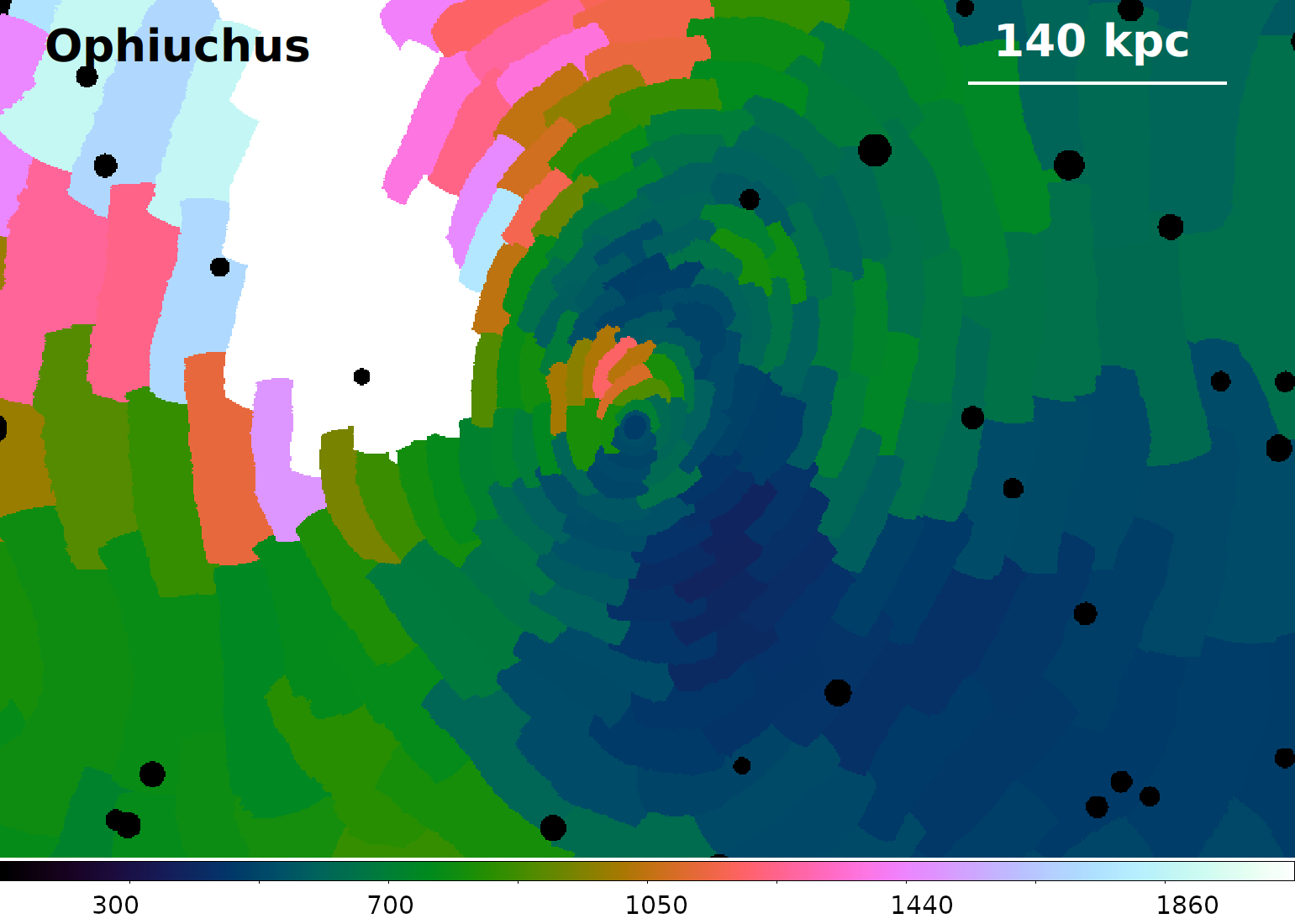} 
\caption{Velocity map errors (km/s) obtained for the Virgo (top panel), Centaurus (middle panel) and Ophiuchus cluster (bottom panel).} \label{fig_contbin_vel_errors} 
\end{figure}

\begin{figure}    
\centering
\includegraphics[width=0.48\textwidth]{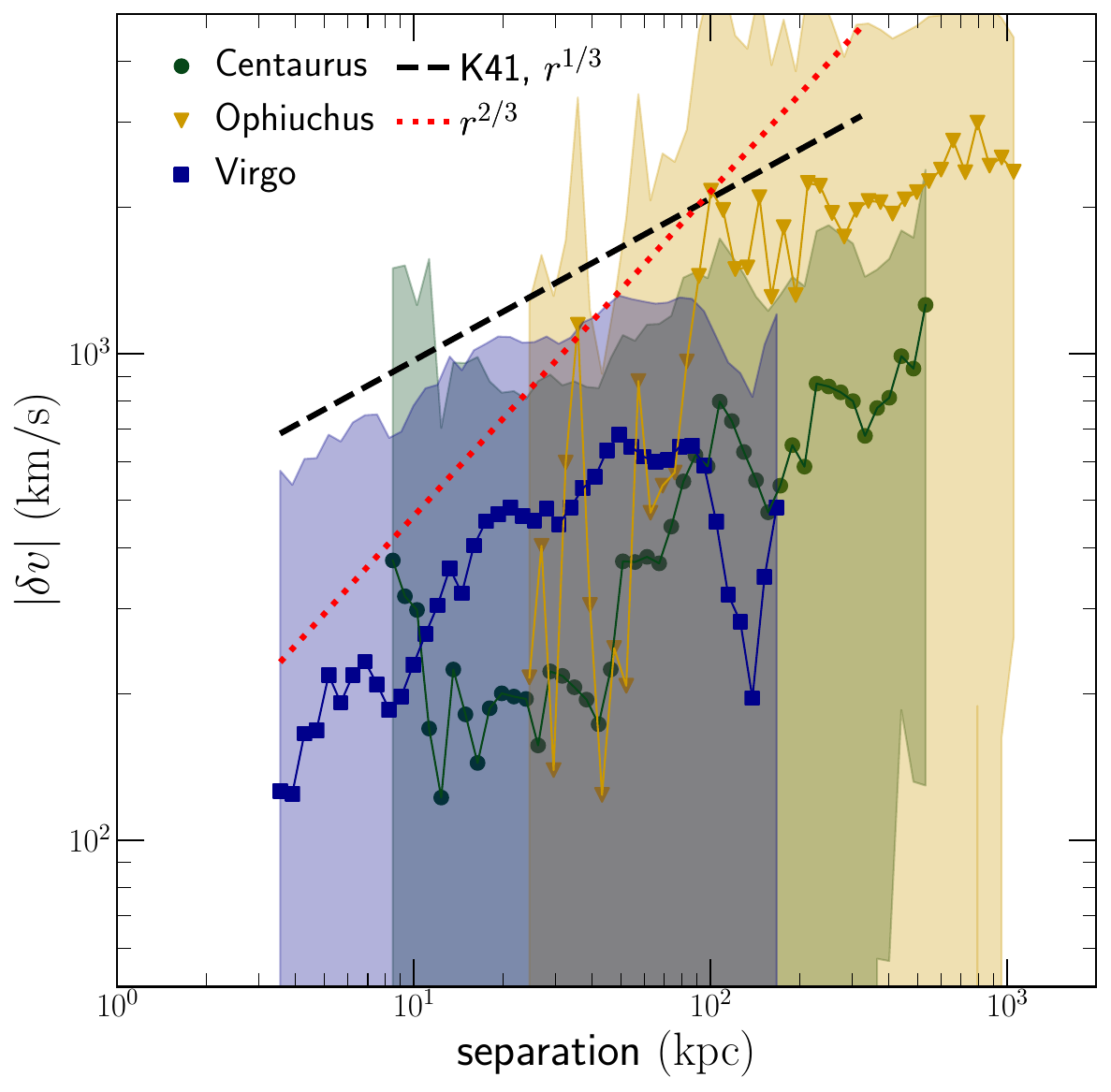}\\
\includegraphics[width=0.48\textwidth]{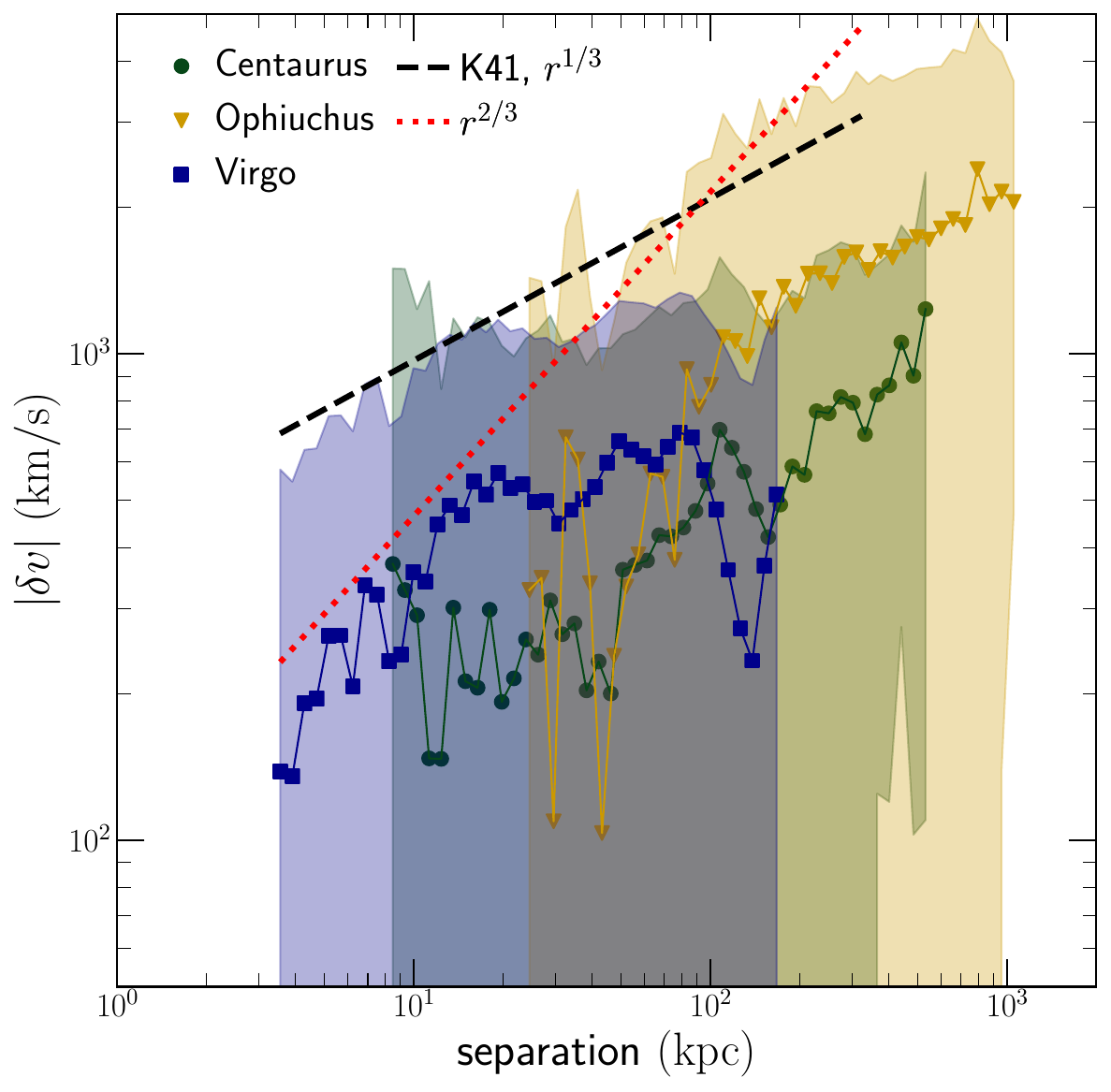}  
\caption{VSFs of the hot ICM obtained for all sources including area weighting (top panel) and error measurements weighting (bottom panel).} \label{fig_vsfs_systematic} 
\end{figure}

\begin{figure*}    
\centering
\includegraphics[width=0.98\textwidth]{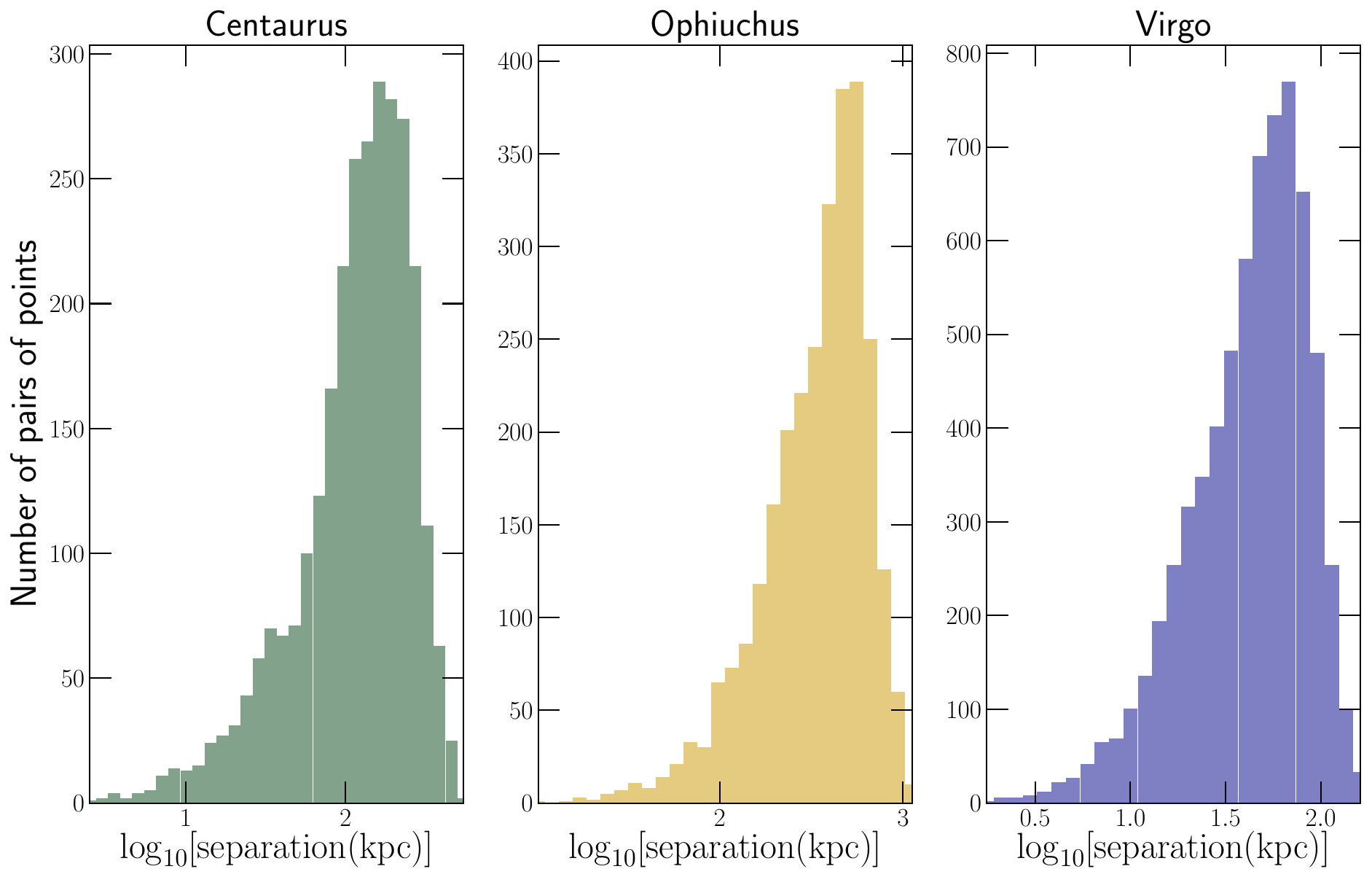} 
\caption{Distribution of pair separations in Centaurus, Ophiuchus and Virgo galaxy clusters.} \label{fig_num_points_per_bin} 
\end{figure*}

The velocity maps obtained for these systems are not equally spaced and therefore there is no pixel-velocity one-to-one relation (see Figure~\ref{fig_contbin_vel}). In order to account for such effects, we have computed the VSF by weighting each region with its area in pixels units. The top panel in Figure~\ref{fig_vsfs_systematic} shows the results. We note that when including the area weighting, the flattening of the Virgo and Centaurus cluster VSFs is less pronounced. We perform a further test on the impact of systematics by weighting the curves including the uncertainties of each velocity measurement (see Figure~\ref{fig_contbin_vel_errors}). The bottom panel in Figure~\ref{fig_vsfs_systematic} shows the VSFs obtained after weighting with the uncertainties. We note that the flattening of the curves is more noticeable. For the Ophiuchus cluster, the flattening shown at large distances in Figure~\ref{fig_vsfs} is no longer present. These results indicate that the area weighting is more sensitive to large-scale variations, while the error weighting is more sensitive on small scale. Figure~\ref{fig_num_points_per_bin} shows the distribution of pair separations in Centaurus, Ophiuchus and Virgo. Pair separations for the Ophiuchus cluster are much larger in comparison with the other sources. It is also clear that our analysis covers intermediate to large spatial scales compared to H$\alpha$ studies.


\subsubsection{Effect of S/N cutoff}\label{SN_cutoff}
\begin{figure*}    
\centering
\includegraphics[width=0.4\textwidth]{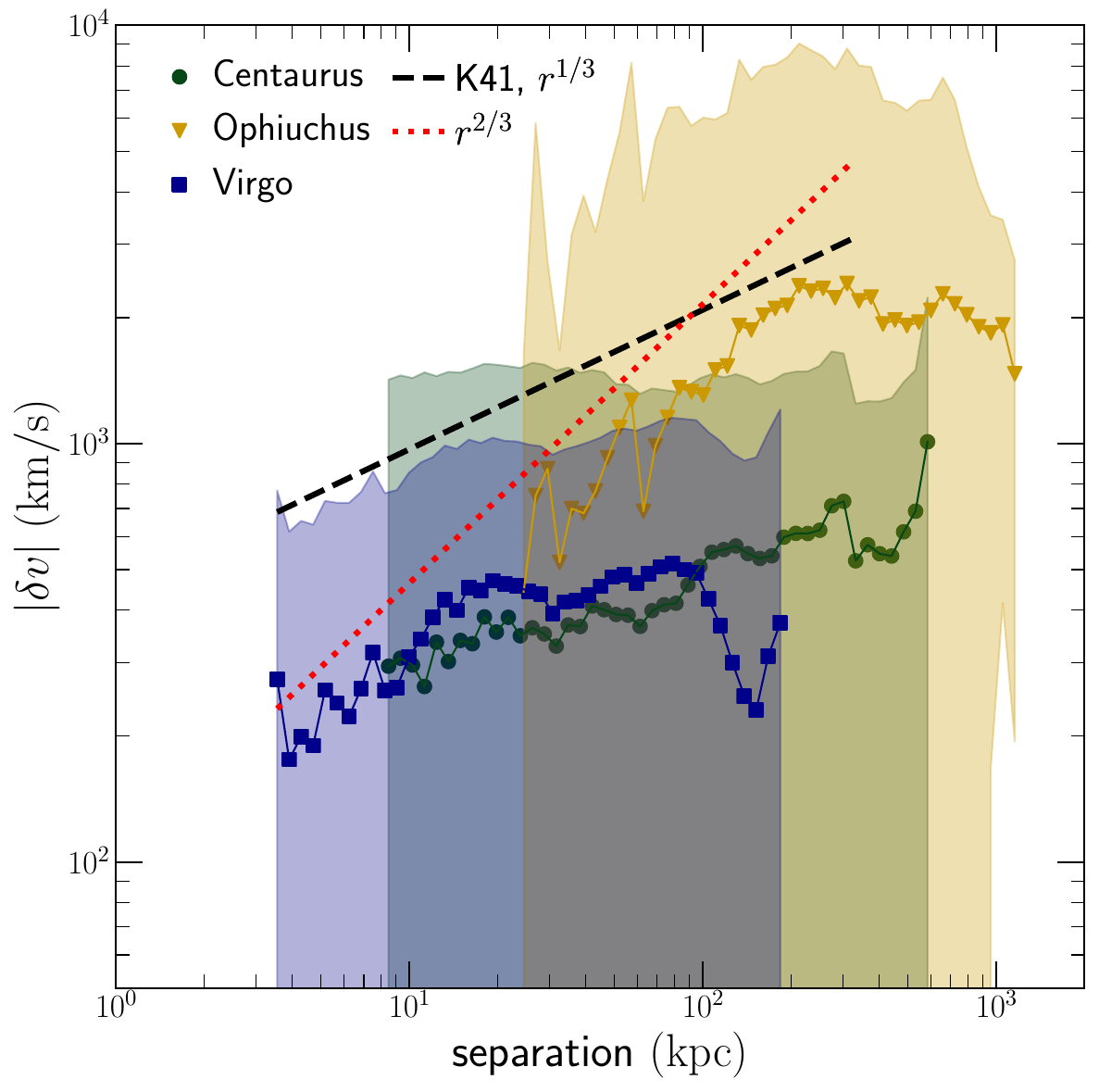}
\includegraphics[width=0.58\textwidth]{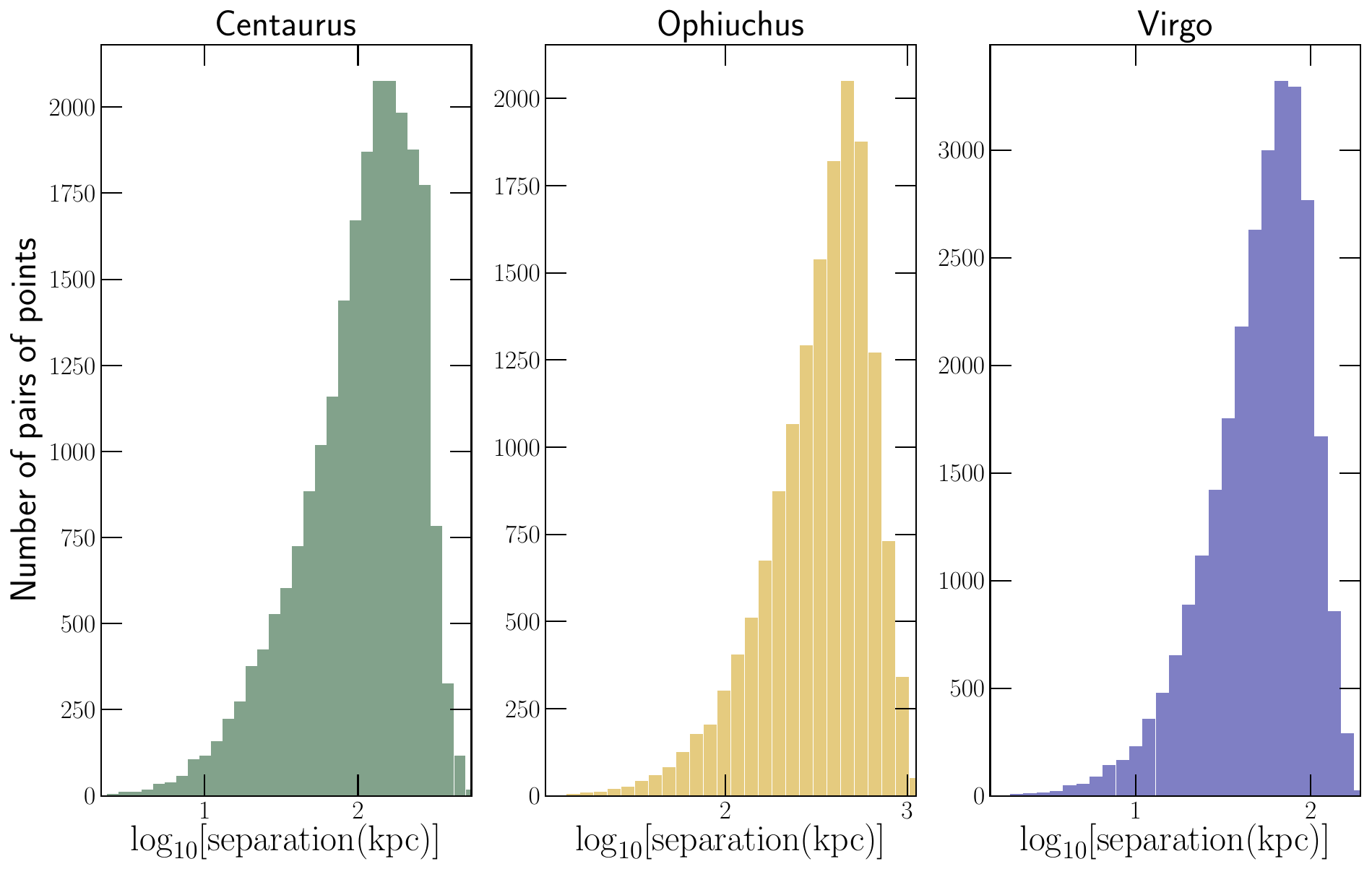}\\
\includegraphics[width=0.4\textwidth]{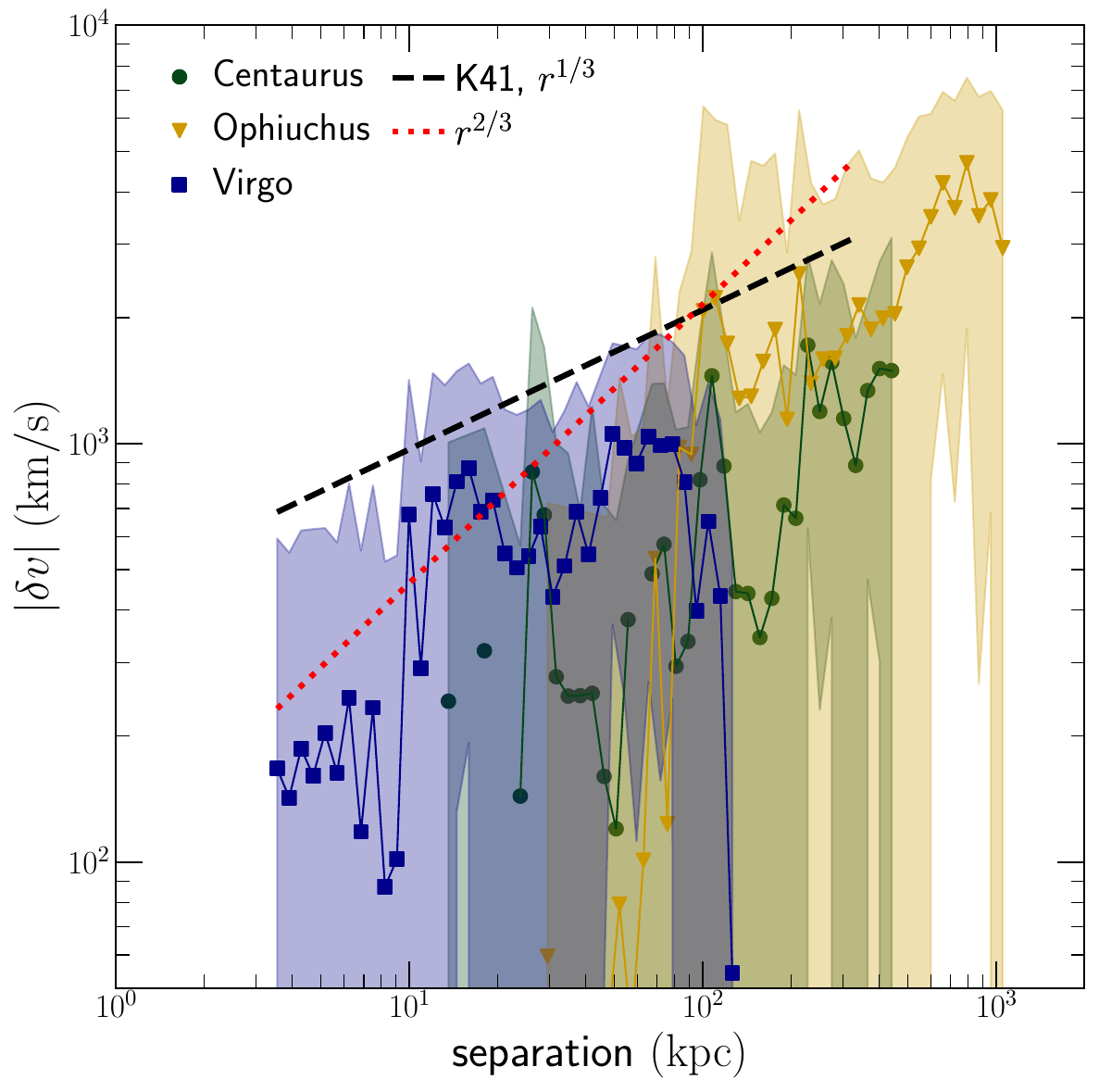}  
\includegraphics[width=0.58\textwidth]{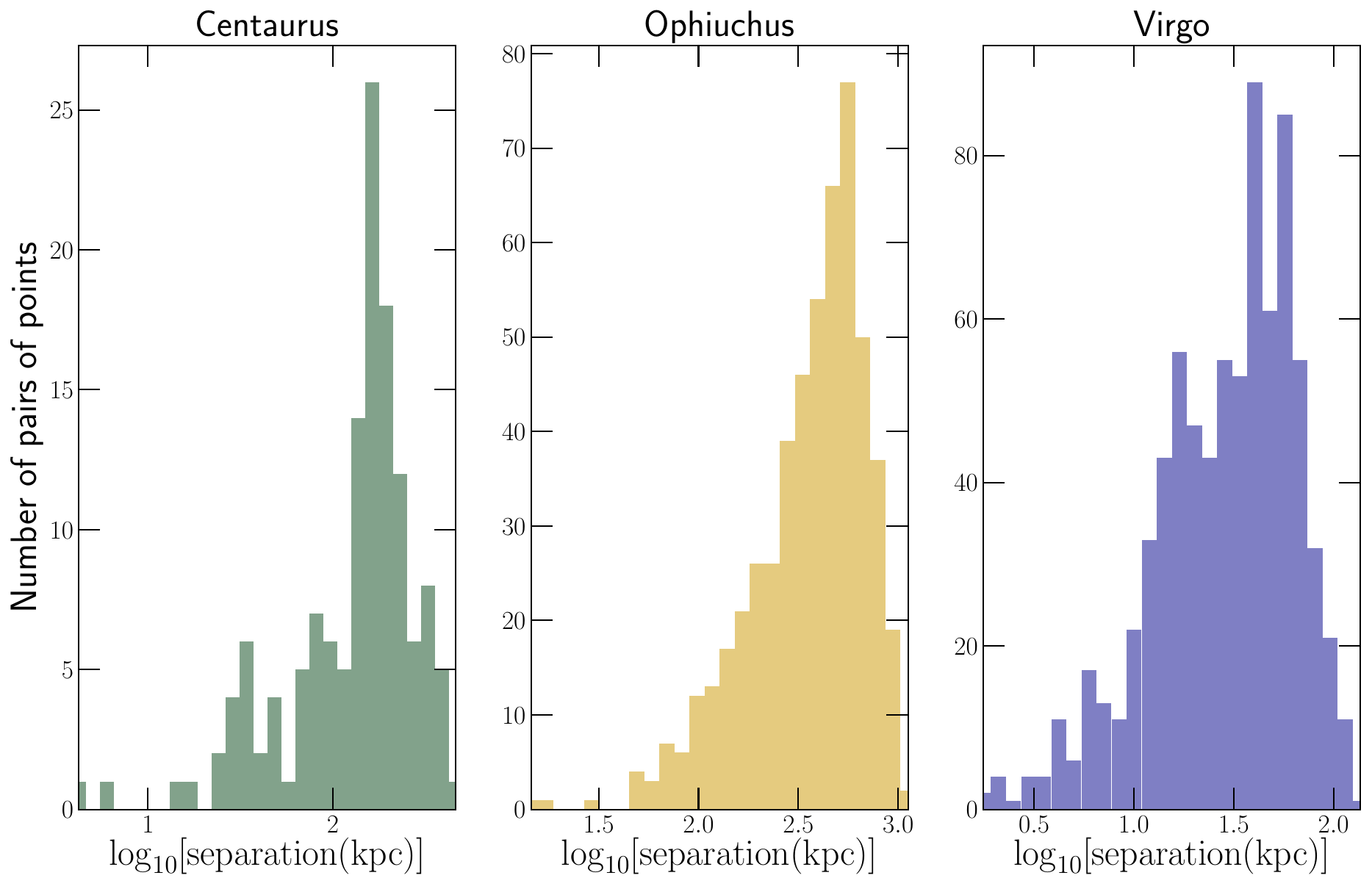}
\caption{VSFs of the hot ICM (left) and number of points considered (right) obtained for all sources including data with $S/N>0.5$ (top panel) and $S/N>1.5$ (bottom panel).} \label{fig_vsfs_sn} 
\end{figure*}

For calculating the $\mathrm{VSF}$s in section \ref{first_vsf}, we have used a signal-to-noise ($S/N$) filter of $1.0$ on the independent velocity dataset. In Fig.~\ref{fig_vsfs_sn} we show the effect of choosing $S/N$ filter to be $0.5$ (upper panel) and $1.5$ (lower panel), respectively. A lower $S/N$ filter gives us a much larger number of pairs of points per radial separation bin and smoother variations in the $\mathrm{VSF}$ with separation. However, these $\mathrm{VSF}$s show larger propagation error (error propagated from $\delta v/v$ measurement). On the other hand, our $\mathrm{VSF}$s with $S/N\geq1.5$ show smaller errorbars but larger scatter and suffer from low number statistics (larger Poisson error), since the number of pairs per separation bin is greatly reduced.

\section{Conclusions and summary}\label{sec_con} 
 
We have analyzed the velocity structure functions (VSFs) of the hot ICM within the Virgo, Centaurus and Ophiuchus clusters of galaxies. This is the first time such velocity structures are measured for the hot gas using direct velocity measurements from X-ray astronomical observations. Line-of-sight velocities were measured using the technique developed by \citet{san20,gat22a,gat22b} to calibrate the absolute energy scale of the {\it XMM-Newton} EPIC-pn detector. Here we briefly summarize our findings.

\begin{enumerate}
\item We made spectral maps of the clusters using the contour binning algorithm. These maps provide velocity measurements for statistically independent regions.
\item We computed the velocity PDFs from the velocity maps. We applied a normality test and found that for all sources the PDF follows a normal distribution, as predicted by simulations. In the case of Ophiuchus, there are hints for a multimodal distribution.
\item We have computed the VSFs for all sources. For the Virgo 
cluster we found a driving scale of the turbulence of $\sim 10-20$~kpc. For the Ophiuchus cluster, the VSF obtained reflects the absence of strong interactions between the ICM and a powerful AGN at such spatial scales. 
\item We have found that the dissipation time is larger than the jet activity cycle, thus indicating that an additional process besides turbulence is required to reach equilibrium. That is, more efficient heating processes are required to reach equilibrium in addition to turbulence. 
\end{enumerate}

\section{Acknowledgements} 
The authors thank Irina Zhuravleva, Yuan Li and Shalini Ganguly for sharing data for Figures~\ref{fig_vsf_virgo_hc} and \ref{fig_vsf_centaurus_hc}. This work was supported by the Deutsche Zentrum f\"ur Luft- und Raumfahrt (DLR) under the Verbundforschung programme (Messung von Schwapp-, Verschmelzungs- und R\"uckkopplungsgeschwindigkeiten in Galaxienhaufen). This work is based on observations obtained with XMM-Newton, an ESA science mission with instruments and contributions directly funded by ESA Member States and NASA. This research was carried out on the High Performance Computing resources of the cobra cluster at the Max Planck Computing and Data Facility (MPCDF) in Garching operated by the Max Planck Society (MPG).
C.F.~acknowledges funding by the Australian Research Council (Future Fellowship FT180100495 and Discovery Projects DP230102280), and the Australia-Germany Joint Research Cooperation Scheme (UA-DAAD). A.L. acknowledges financial support from the European Research Council (ERC) Consolidator Grant under the European Union's Horizon 2020 research and innovation programme (grant agreement CoG DarkQuest No 101002585).

\subsection*{Data availability}
The observations analyzed in this article are available in the {\it XMM-Newton} Science Archive (XSA\footnote{\url{http://xmm.esac.esa.int/xsa/}}).
 
\bibliographystyle{mnras}
\bibliography{my-references}  

\end{document}